\documentclass[preprint,aps,showpacs]{revtex4}
\usepackage{epsf, epsfig, graphics}

\begin{document}
\draft
\title{Adhesion-induced phase separation of multiple species of 
       membrane junctions}
\author{Hsuan-Yi Chen}
\affiliation{Department of Physics and Center for Complex Systems, 
National Central University, \\ 
Chungli, 32054 \\
Taiwan}

\date{\today}

\begin{abstract}
  A theory is presented for the membrane junction separation induced by the
adhesion between two biomimetic membranes that contain two different types of 
anchored junctions (receptor/ligand complexes).  
The analysis shows that several mechanisms contribute to the membrane 
junction separation.  These mechanisms include 
(i) the height difference between type-1 and type-2 junctions is the 
main factor which drives the junction separation,  
(ii) when type-1 and type-2 junctions have different rigidities against
stretch and compression, 
the ``softer'' junctions are the ``favored'' species, and the aggregation of
the softer junction can occur,
(iii) the elasticity of the membranes mediates a non-local interaction
between the junctions, 
(iv) the thermally activated shape fluctuations of the 
membranes also contribute to the junction separation by inducing another
non-local interaction between the junctions and renormalizing the 
binding energy of the junctions.
The combined effect of these mechanisms is that when junction separation 
occurs, the system separates into two domains with different relative and 
total junction densities.
\end{abstract}

\maketitle
\section{Introduction}
\label{sec:intro}
  Adhesion of membranes is responsible for cell adhesion which plays an
important role in embryological development, immune response, and the
pathology of tumors~\cite{ref:alberts_book}.  
In many cases, membrane adhesion in biological systems is mediated by the 
specific attractive interactions between complementary pairs of ligands 
and receptors which are anchored in the membranes.~\cite{ref:lipowsky_book_95}
At the same time, the adhesion between 
multi-component bio-membranes or biomimetic
membranes is also intimately related to domain formation.~\cite{ref:sackmann_BJ_97,ref:sackmann_EPL_97,ref:sackmann_PRE_00,ref:lipowsky_PRE_00,ref:lipowsky_PRE_01,ref:andelman_EPE_00,ref:bruinsma_PRL_95,ref:andelman_EPE_02,ref:lipowsky_EPL_02}  
When the 
membrane adhesion is mediated by the specific lock-and-key type of bonds
between the anchored ligands and receptors, i.e., junctions, 
adhesion-induced lateral phase separations have been observed in many 
experiments in biomimetic
systems~\cite{ref:sackmann_BJ_97,ref:sackmann_EPL_97,ref:sackmann_PRE_00}.
Theoretical models and Monte Carlo 
simulations~\cite{ref:lipowsky_PRE_00,ref:lipowsky_PRE_01,ref:andelman_EPE_00,ref:bruinsma_PRL_95,ref:andelman_EPE_02,ref:lipowsky_EPL_02}
have also shown similar phase separation behavior in various systems.  

  So far, studies on adhesion-induced lateral phase separation have focused
on the case when the system has a single type of junctions.
The presence of the glycol proteins anchored in the membranes
(i.e., repellers),
and the interplay between generic interactions (for example, Van der
Waals, or electrostatic interactions) and specific ligand/receptor
interactions are believed to enhance this phase separation.  
However, in biological systems membrane adhesion are often mediated by more 
than one type of
junctions, and the adhesion-induced junction separation are believed to play
an important role in some biological processes.  For example, a key event 
governing a mature immune response when T lymphocytes interact  with 
antigen-present cells is the formation of 
immunological synapses. An immunological synapse is a patch of membrane
adhesion region between a T cell and an antigen-present cell, 
where the TCR/MHC-peptide complexes aggregate in the center with
a LFA-1/ICAM-1 complexes rich region surrounds 
it.~\cite{ref:groves_pnas_01,ref:exp,ref:Burroughs_BPJ_02}  
Since a complete understanding of the physical mechanism behind this type of 
adhesion-induced multi-species membrane junction separation is still 
unavailable, in the present work I develop a theoretical model to study the
equilibrium properties of such systems.

This article is organized as follows.
In section~\ref{sec:model}, I discuss a coarse grained
model for the adhesion of two membranes due to the formation
of two types of junctions.  To concentrate on the effect of the differences
between type-1 and type-2 junctions, the glycocalyx and the generic 
interactions between the membranes are not considered in this model.
Furthermore, I assume that the membranes are bound to each other due to the
formation of the membrane junctions.  Hence I will not discuss another 
interesting problem of the unbinding transition. 
An approximate solution of this model which neglects the fluctuations of 
membrane-membrane distance (the ``hard membrane'' solution) is studied in
section~\ref{sec:hard}.
This simplified solution already 
reveals several mechanisms which are important to the phase behavior of
the system.  For example, when type-1 and type-2 junctions have the same 
rigidity, membrane adhesion can induce a junction separation which is
driven by the height difference of the junctions.  In this situation the
membranes separate into a type-1-junction-rich domain and a 
type-2-junction-rich domain.
On the other hand, 
when the junctions have different rigidities and the height difference is
not very large, membrane adhesion can
induce an aggregation of the ``softer'' junctions, i.e., the membranes
can separate into two domains which are both rich in the softer junctions.  
Thus, in general situations both
mechanisms contribute to the adhesion-induced junction separation.  When
phase separation occurs, the system separates into two domains with different
membrane-membrane distance because of the height mismatch of the junctions, 
and the total number of the softer junctions in the system is greater than the
total number of the stiffer junctions.   

The fact that the hard membrane solution assumes that the membrane-membrane 
distance is a constant has its drawback, too.  An apparent artifact of the
hard membrane solution is that
when junction separation occurs, the total junction density, i.e., 
$\phi _1+\phi _2$ ($\phi _{\alpha}$ is the density of type-$\alpha $ 
junctions), is the same in domains which have different values of
$(\phi _1 -\phi _2)/(\phi _1+\phi _2)$ (the relative densities of the
junctions)!
This artifact also shows that, in order to acquire a complete physical 
picture of the junction separation, 
the effect of non-constant membrane-membrane distance,
and the thermally activated fluctuations of the membranes 
and junction densities should be taken into account.    
Therefore in section ~\ref{sec:fluc} I study the effects of these fluctuations.
The fluctuation analysis shows that, first, the thermally activated membrane 
fluctuations renormalize the chemical potentials of 
the junctions, and effectively reduce the binding energies of the junctions.  
This chemical potential renormalization is less significant for the softer 
junctions because they allow the membranes more freedom to move.
Second, the fluctuation analysis also reveals nonlocal interactions between
the junctions which are mediated by the membrane elasticity and thermally 
activated fluctuations of the junction densities and membrane-membrane distance
.  These interactions are not included in the simple physical picture provided
by the hard membrane solution.
As a result of these effects, when junction separation occurs, 
domains with different values of $(\phi _1 -\phi _2)/(\phi _1+\phi _2)$
also have different values of $\phi _1 + \phi _2$.  
The fluctuation analysis also shows that, when the hard membrane solution
of the junction densities are small, or when the junctions are very short
or very soft, the membrane fluctuations are sufficiently large such that 
the present analysis cannot
provide the complete physical picture for the system.  This criterion shows 
under what conditions one needs a numerical simulation of the model to
provide a better picture of the physics in this system.   
Section~\ref{sec:summary} summarizes this work.  The Appendix discusses
the details of the fluctuation analysis around the hard membrane solution.

\section{The model}
\label{sec:model}
  To focus on the physics of adhesion-induced phase separation, I will 
not discuss the binding/unbinding transition but only consider the case 
when the membranes are bound to each other due to the presence of the 
junctions.  The system is shown schematically in Fig.~1.
The heights of the membranes measured from the reference plane (i.e., the 
$xy$-plane) are denoted as $z_1({\bf r})$ and $z_2({\bf r})$, respectively, 
where ${\bf r}=(x,y)$ is a two-dimensional planar vector.  
There are two types of anchored receptors in membrane 1, and
two types of anchored ligands in membrane 2.  Type-$\alpha$ receptors
($\alpha$ is $1$ or $2$) form specific lock-and-key complexes with 
type-$\alpha$ ligands, these are the junctions which mediate the 
membrane adhesion.
The density of type-$\alpha$ junctions at ${\bf r}$ is 
$\phi _{\alpha}({\bf r})$, 
and the densities of free type-$\alpha$ receptors and ligands at 
${\bf r}$ are denoted by 
$\psi_{R\alpha}({\bf r})$ and $\psi_{L\alpha}({\bf r})$, respectively.
The binding energy of a type-$\alpha$ junction is denoted by 
$E_{B\alpha}$. 
    
   The effective Hamiltonian of the system can be written as    
\begin{eqnarray}
H = \int d^2 r \left\{ \frac{\kappa}{2} \left(\nabla^2 h({\bf r})\right)^2  
    + \frac{\gamma}{2} \left(\nabla h({\bf r})\right)^2
    + \sum _{\alpha = 1}^{2} \frac{\lambda _{\alpha}}{2} \phi_{\alpha}({\bf r})
                            \left( h({\bf r}) - h_{\alpha} \right)^2
    - \sum _{\alpha = 1}^{2} \phi _{\alpha} E_{B\alpha}
\right\}.
\label{eq:hamiltonian_1}
\end{eqnarray}
The energy unit is chosen to be $k_BT$. 
Here $h({\bf r}) = z_1({\bf r})-z_2({\bf r})$ is the membrane-membrane distance
at ${\bf r}$.  
The first and second terms on the right hand side are the bending 
elastic energy and surface tension of the membranes.  
$\kappa$ is related to the bending moduli of the membranes by 
$\kappa = \kappa _1 \kappa _2/(\kappa _1+\kappa _2)$~\cite{ref:lipowsky_PRL_96}
, and $\gamma$ is related to the surface tension of the membranes by
$\gamma = \gamma _1 \gamma _2/(\gamma _1+\gamma _2)$~\cite{ref:lipowsky_PRL_96}.  In this simple model it is assumed that $\kappa$ and $\gamma$ are 
independent of the densities of the receptors and ligands anchored in the 
membranes.  
I also assume that in the 
presence of a type-$\alpha$ junction, the interaction energy between the 
membranes acquires a minimum at $h=h_{\alpha}$ (the natural height of a 
type-$\alpha$ junction), and the coupling term 
$\sum _{\alpha = 1}^{2} \frac{\lambda _{\alpha}}{2} \phi_{\alpha}({\bf r})
\left( h({\bf r}) - h_{\alpha} \right)^2$
comes from the Taylor expansion around this minimum.  
Here $\lambda _{\alpha}$ is the rigidity of a
type-$\alpha$ junction against stretch or compression.  
The last term on the right hand side is the binding energy between the
receptors and the ligands.
To focus on the effect of 
adhesion-induced interactions, I have neglected all the direct interactions 
between the junctions, receptors, and ligands.  The nonspecific interactions
between the membranes are also neglected.
For simplicity, from now on I further choose the unit length in the 
$xy$-plane to be 
$\sqrt{a}$, where $a$ is the in-plane size of an inclusion,   
and the unit length in the $z$-direction is chosen to be 
$\sqrt{a/\kappa} \equiv l_0$.  Thus the Hamiltonian of 
the system can be expressed in the non-dimensional form,
\begin{eqnarray}
H = \int d^2 r \left\{ \frac{1}{2} \left(\nabla^2 h({\bf r})\right)^2  
    + \frac{\Gamma}{2} \left(\nabla h({\bf r})\right)^2
    + \sum _{\alpha = 1}^{2} \frac{\Lambda _{\alpha}}{2} \phi_{\alpha}({\bf r})
                            \left( h({\bf r}) - h_{\alpha} \right)^2
    - \sum _{\alpha = 1}^{2} \phi _{\alpha} E_{B\alpha} \right\},
\end{eqnarray}
where $\Gamma = \gamma l_0^2$ is the dimensionless surface tension,
$\Lambda _{\alpha}=\lambda _{\alpha} l_0^2$ is the dimensionless junction
rigidity, and all in-plane lengths and heights are scaled by $\sqrt{a}$ and 
$\sqrt{a/\kappa} \equiv l_0$, respectively.

The effective interaction free energy between the junctions due 
to the membrane-junction coupling is obtained by integrating over $h({\bf r})$,
\begin{eqnarray}
   F_c[\phi_{\alpha}] = -\ln \left( \int D[h] e^{-H[h,\phi_{\alpha}]} \right).
\label{eq:fc_exact}
\end{eqnarray}
Thus in the spirit of density functional theory, the total free energy of the
system is provided by 
\begin{eqnarray}
 F = F_c + F_s,
\label{eq:free_energy}
\end{eqnarray}
where
\begin{eqnarray}
F_s = \sum_{\alpha =1}^{2}\int d^2r \left( 
      \phi_{\alpha}({\bf r}) ( \ln \phi_{\alpha} -1)
    + \psi_{R\alpha}({\bf r}) ( \ln \psi_{R\alpha} -1)
    + \psi_{L\alpha}({\bf r}) ( \ln \psi_{L\alpha} -1)
      \right) 
\end{eqnarray}
is the contribution from the entropy of the junctions, receptors and ligands.
Here I have assumed that $\phi _{\alpha} \ll 1$, 
$\psi _{R \alpha} \ll 1$, and $\psi _{L \alpha} \ll 1$.
In principle, once $F_c$ is calculated, the equilibrium distribution of 
the junction density is determined by minimizing the total free energy
of the system under the constraint that the total numbers of the receptors
and ligands in the system are fixed, i.e, 
\begin{eqnarray}
  \int d^2 r \left\{ \phi _{\alpha}({\bf r}) + \psi _{R \alpha}({\bf r})
             \right\} 
&=& N_{R \alpha},
\nonumber
\\
  \int d^2 r \left\{ \phi _{\alpha}({\bf r}) + \psi _{L \alpha}({\bf r}) 
             \right\}
&=& N_{L \alpha},
\label{eq:cons}
\end{eqnarray}
here $N_{R \alpha}$ and $N_{L \alpha}$ are the total number of type-$\alpha$ 
receptors and ligands in each membranes when the membranes are completely 
detached.  

\section{``Hard membrane'' solution}
\label{sec:hard}
   Since the integral in Eq.~(\ref{eq:fc_exact}) cannot be carried 
out exactly, in this section I discuss an approximate solution in which  
$\phi _{\alpha}$ and $h({\bf r})$ are independent of ${\bf r}$.
In this approximation, $F_c$ can be easily calculated by looking for the
saddle point in the integrand.  
This is equivalent to neglecting the fluctuations
of the membrane-membrane distance, therefore 
I call this mean-field approximate 
solution the ``hard membrane'' solution.  To simplify the notation, I define 
$\Lambda _{\pm} = \Lambda _1 \pm \Lambda _2$,
$\phi _{\pm} = (\phi _1 \pm \phi _2)/2$, and let
$h_1 = h_0 -\triangle _h$, $h_2 = h_0 + \triangle _h$. Thus the hard membrane
solution of the membrane-membrane distance can be expressed by 
\begin{eqnarray}
h &=& h_0 - \frac{\Lambda _{-} \phi _{+} + \Lambda _{+} \phi _{-}}{
                  \Lambda _{+} \phi _{+} + \Lambda _{-} \phi _{-}} \triangle _h
\nonumber \\
  &\equiv& h_0 - l_M.
\end{eqnarray}
Notice that $l_M$ depends on the junction densities.
After substituting $h$ back to the Hamiltonian, the effective interaction
free energy between the junctions, $F_c$, can be expressed by its saddle-point
value
\begin{eqnarray}
F_c = \int d^2 r \left\{ 
      \frac{\Lambda _{+} \phi _{+} + \Lambda _{-} \phi _{-}}{2}
        \left( l_M^2 + \triangle _h^2  \right)
      + \left ( \Lambda _{-} \phi _{+} + \Lambda _{+} \phi _{-} \right)
        l_M \triangle _h 
      - \left( E_{+} \phi _{+} + E_{-} \phi _{-} \right)  \right\},
\label{eq:fcmft}
\end{eqnarray}
where $E_{\pm} = E_{B1} \pm E_{B2}$.  It is clear that there is an
interaction between the junctions due to the membrane adhesion. 
To minimize the total free energy under the constraints in Eq.~(\ref{eq:cons}),
it is convenient to work in the grand canonical ensemble and define the free 
energy $G$ of the system under constant chemical potentials, 
\begin{eqnarray}
G =  F_c + F_s  
   - \sum _{\alpha}\mu _{R\alpha}\int d^2r(\phi _{\alpha} + \psi _{R\alpha})
   - \sum _{\alpha}\mu _{L\alpha}\int d^2r(\phi _{\alpha} + \psi _{L\alpha}).
\end{eqnarray}
The chemical potentials, $\mu _{R\alpha}$, $\mu _{L\alpha}$ are determined by
fixing the total number of receptors and ligands in the system.  However,
for convenience I will proceed the discussion in the grand canonical ensemble.
After some straightforward algebra, $G$ is expressed as
\begin{eqnarray}
G = \int d^2r \ \phi _{+} \left\{ g(\phi )   
                                + 2 \left( \ln \phi _{+} -1 \right) 
                                - \mu _{+} \right\}
    + G_{\psi},
\label{eqn:MFT}
\end{eqnarray}
where
\begin{eqnarray}
g(\phi ) &=& -\frac{\triangle _h^2}{2}\Lambda _{+} 
        \frac{(\lambda +\phi)^2}{1+\lambda \phi}  
       +\left( 1+\phi \right) \ln \left( 1+ \phi \right)
       +\left( 1-\phi \right) \ln \left( 1- \phi \right)
       -\mu _{-} \phi \nonumber \\
         &\equiv & f(\phi ) -\mu _{-} \phi ,
\end{eqnarray}
$\lambda \equiv \Lambda _{-} / \Lambda _{+}$, 
      $\phi \equiv \phi _{-} / \phi _{+}$, 
and $\mu _{\pm} = (\mu _{R1} + \mu _{L1}) \pm 
                  (\mu _{R2} + \mu _{L2}) + E_{B1} \pm E_{B2} 
     - \frac{\triangle _h^2}{2} \Lambda _{\pm}$.
$G_{\psi}$ includes terms which only depend on $\psi _{R \alpha}$ and
$\psi _{L \alpha}$, they are decoupled from the other terms, hence
from now on I neglect $G_{\psi }$. 
From Eq.~(\ref{eqn:MFT}), it is clear that in the hard membrane solution 
the phase behavior of the junctions is governed
by $\triangle _h^2 \Lambda _{+}$, $\lambda$, and $\mu _{\pm}$.
Minimizing $g(\phi)$ leads to the equilibrium value  of $\phi $, and later 
I will show that there can be a phase separation in $\phi $.  
On the other hand, from Eq.~(\ref{eqn:MFT}), $\phi _{+}$ satisfies
\begin{eqnarray}
\phi _{+} = \exp \left( \frac{1}{2} \mu _{+} - \frac{1}{2} g(\phi) \right).
\label{eq:phipM}
\end{eqnarray}
Because in equilibrium $\phi $ is determined by minimizing $g(\phi )$, 
$g(\phi)$ takes single value even when there is a phase separation
in $\phi$.  Therefore in the hard membrane solution $\phi_+$ is  
single-valued even when the system separates into two domains with 
different values of $\phi$!
In the next section I will show
that, when the effects of fluctuations around the hard membrane solution 
are taken into account, the analysis reveals a renormalization of the 
binding energy of the junctions and (nonlocal) interactions 
between the junctions which are mediated by the membrane elasticities and 
thermally activated membrane fluctuations. 
As a result, the true equilibrium solution of $\phi _{+}$ is not 
single-valued in the regime where junction separation happens.
Thus, the fact that $\phi _{+}$ is single-valued in the hard membrane 
solution is an artifact of the approximation which assumes 
constant junction densities and membrane-membrane distance.

Now I discuss the hard membrane solution of $\phi$.  To emphasize different
roles played by $\triangle _h^2 \Lambda _{+}$ and $\lambda$, I begin the 
discussion with the special case when $\lambda = 0$, 
i.e., when both types of junctions
have the same rigidities.  In this case the important parameter of the theory 
is  $\triangle _h^2 \Lambda _{+}$, and $g(\phi)$ has a very simple form 
\begin{eqnarray}
 g(\phi ) = -\frac{\triangle _h^2 \Lambda _{+}}{2} \phi ^2
             +\left( 1+\phi \right) \ln \left( 1+ \phi \right)
             +\left( 1-\phi \right) \ln \left( 1- \phi \right)
             -\mu _{-} \phi .
\end{eqnarray}
This form is exactly the same as the Flory-Huggins theory for
binary mixtures~\cite{ref:doi_book}, where phase separation occurs when
 $ \triangle _h^2 \Lambda _{+} >2$ and the phase coexistence curve is a 
straight line at $\mu_-=0$.  This phase coexistence curve ends at a critical 
point $\mu _{-} =0$, $\triangle _h^2 \Lambda _{+}=2$.
The physics in this special case $\lambda =0$ is clear: the difference in 
junction height drives a junction separation, and this separation only occurs
when the factor  $ \triangle _h^2 \Lambda _{+}$, a combination of
junction height difference and junction rigidity, is sufficiently large. 
On the phase coexistence curve, the system separates into $\phi _1$-rich
and $\phi _2$-rich domains, and the system is symmetric under 
$\phi \rightarrow -\phi$.

Next I discuss the more general case $\lambda \neq 0$, i.e., the junctions
have different rigidities.  Fig.~2
shows the shape of $g(\phi)$ with different values of $\mu _{-}$ when 
$\lambda=0.2$ and $\triangle _h^2 \Lambda _{+} = 1.998$.  
Notice that this is the case when $\triangle _h^2 \Lambda _{+} <2$, 
i.e., there is no junction separation if $\lambda =0$.
Nevertheless, Fig.~2
clearly shows that $g(\phi)$ has 
two local minimum, both at negative $\phi$, 
and phase coexistence occurs when $\mu _{-} \approx -0.4045$. 
Since this is the case when 
$\lambda >0$, i.e., type-2 junctions are ``softer'' than type-1 junctions, 
double minimum at $\phi = \phi _1 - \phi _2 <0$ means that 
the softer junctions tend to aggregate, when phase coexistence occurs a
domain with high $\phi _2 - \phi _1$ coexists with a domain with small
$\phi _2 - \phi _1$.  The density of type-2 junctions is higher than the
density of type-1 junctions in both domains.

Another effect of nonzero $\lambda$  can be seen in Fig.~3
, where $g(\phi)$ for different values of $\lambda$ is shown at 
$\triangle _h^2 \Lambda _{+}=2.04>2$.  It shows that $g(\phi)$ is 
symmetric in $\phi$ when $\lambda=0$ but asymmetric in $\phi$
for nonzero $\lambda$, i.e., the symmetry under $\phi \rightarrow - \phi$
no longer exists when the junctions have different rigidities.  
Comparing to $\lambda =0$ case, in the case when $\lambda >0$, 
the minima of $g(\phi)$ are shifted towards smaller $\phi$ values, i.e.,
the softer junctions are easier to be formed.  
Notice that different from the example in Fig.~2, 
in Fig.~3 when phase coexistence occurs the membranes separate into 
$\phi _1$-rich and $\phi _2$-rich domains, but the
softer junctions (in this case type-2 junctions) are the ``favored'' species,
i.e., the total number of the softer junctions in the system is greater than
the total number of the stiffer junctions. 
From these two examples of nonzero $\lambda$, 
I  conclude that in general the experimentally observed junction 
separation induced by membrane adhesion is actually a result of the combined 
effect of the aggregation of softer junctions and the separation of the 
junctions due to the mismatch of junction heights.  

In the neighborhood of 
$\triangle _h^2 \Lambda _+ =2$, $\lambda =0$, 
the equilibrium value of $\phi$ is small compared to unity, therefore 
the phase diagram of the system in this regime 
can be studied by expanding $g(\phi)$ around $\phi =0$,
\begin{eqnarray}
g(\phi) &=&  r_2 \phi ^2 +r_3 \phi ^3 + r_4 \phi ^4 - \tilde{\mu} _-\phi
             + const.+\mathcal{O}(\phi ^5), \\
r_2 &=& 1- \frac{\triangle _h^2}{2} \Lambda _+
                   \left( 1-\lambda ^2 \right)^2,  \nonumber \\
r_3 &=& \frac{\triangle _h^2}{2}\Lambda _+ \lambda 
                   \left(1-\lambda^2 \right)^2,  \nonumber \\
r_4 &=& \frac{1}{6} 
              -\frac{\triangle _h^2}{2}\Lambda _+ \lambda ^2
 		   \left( 1-\lambda^2 \right)^2, \nonumber \\
\tilde{\mu}_- &=& \mu _- + \frac{ \triangle _h^2}{2} \Lambda _+ \lambda
	 	   \left(2- \lambda ^2\right), \nonumber \\
const. &=& \frac{ \triangle _h^2}{2} \Lambda _+ \lambda ^2,
\end{eqnarray} 
and $\mathcal{O}(\phi ^5)$ is the contribution from terms of order
$\phi ^5$ and higher.
The phase diagram in the neighborhood of 
$\triangle _h^2 \Lambda _+ =2$, $\lambda =0$
is plotted schematically in Fig.~4.
where the phase coexistence curve for $\lambda =0$
ends at a critical point $\triangle _h^2 \Lambda _+ =2, \ \mu_-=0$, and
the end points of the phase coexistence curves for $\lambda \neq 0$
occurs at the triple root of $\partial g/\partial \phi =0$.  
Straightforward calculation leads to the position of the end points
of the phase coexistence curves at
\begin{eqnarray}
\triangle _h^2 \Lambda _+ &=&2(1-9 \lambda ^2/4)+\mathcal{O}(\lambda ^4),
\nonumber \\ 
\mu _- &=& -2 \lambda + \mathcal{O}(\lambda ^3).  
\end{eqnarray}
This shows how the smallest value of $\triangle _h^2 \Lambda _+$ above which
phase separation can occur decreases as the difference of junction rigidities
increases.  The phase coexistence
curves move towards the $\lambda=0$ phase boundary as the value of 
$\frac{\triangle _h^2}{2} \Lambda _+$ increases.  This is because as 
$\frac{\triangle _h^2}{2} \Lambda _+$ increases, the effect of junction
height mismatch becomes more important, and the difference in the junction
rigidities becomes less important.
 
Although the hard membrane solution is a very simplified analysis of the
model, it nevertheless reveals interesting physics of the junction separation
due to membrane adhesion.  First of all, the height difference between 
different types of junctions is not the only factor which is important for
the junction distribution.  It is only when the rigidities of both types 
of junctions are the same that the difference in the junction height is the
most important factor in the junction separation.  If the system consists of 
junctions with different rigidities, the softer junctions are more favored,
and the junction separation is a result of the interplay between the 
aggregation of the softer junctions and the separation of the junctions due
to the height difference.  Therefore the smallest value of 
$\triangle _h^2 \Lambda _+$ above which phase separation can occur is 
smaller for systems with larger $|\lambda|$.
However, the approximations in the hard membrane
solution lead to the surprising result that 
$\phi _+$ is the same in both domains when junction separation occurs.  
This indicates that analysis which includes the elasticity of the membranes
and the thermal fluctuations around the hard 
membrane solution is very much needed for a full understanding of the nature 
of the membrane adhesion-induced interactions between the junctions.

\section{Beyond ``hard membrane'' solution}
\label{sec:fluc}
As mentioned in the previous section, the hard membrane solution neglects
the effects of nonuniform membrane-membrane distance and junction densities,
therefore fails to take the effects of membrane-mediated nonlocal interactions
between the junctions into account.  The result is reflected in the unrealistic
solution of single valued $\phi _+$ in both domains when phase coexistence 
occurs. To study these membrane-mediated effects, in this section 
I include the fluctuations of 
membrane-membrane distance and junction densities by expanding the free energy 
of the system around the hard membrane solution.  In the following I denote 
the true membrane-membrane distance as
\begin{eqnarray}
   h({\bf r}) = h_0 + l_M + \delta l({\bf r})\equiv h_M + \delta l({\bf r}),
\end{eqnarray}
and the densities of the junctions are expressed by 
\begin{eqnarray}
   \phi _{\alpha} = \phi _{\alpha M} + \delta \phi ({\bf r}).
\end{eqnarray}
Here $\delta l$, $\delta \phi _{\alpha}$ are the deviations of the true 
values of $h$ and $\phi _{\alpha}$ from their hard membrane solutions,
$\phi _{\alpha M}$ and $h_M$ are the hard membrane solution of 
$\phi _ \alpha({\bf r})$ and $h({\bf r})$, respectively.  
In this expansion, the coarse grained Hamiltonian can be expressed as 
\begin{eqnarray}
H = H _{M} + H _{0} + H_{1} + H_{\phi},
\end{eqnarray}
where $H_{M}$ is $H(h_M, \phi _{1M}, \phi _{2M})$, 
\begin{eqnarray}
H_{0} = \int d^2 r \left\{ \frac{1}{2} (\nabla ^2 \delta l)^2 
                  +\frac{1}{2} \Gamma (\nabla \delta l)^2
  +\left[ l_M (\Lambda _1 \delta \phi _{1} + \Lambda _2 \delta \phi _{2})
    + \triangle _h (\Lambda _1 \delta \phi _{1} - \Lambda _2 \delta \phi _{2})
   \right] \delta l        \right\}
\end{eqnarray}                       
includes terms which are bilinear in $\delta l$ and 
$\delta \phi _{\alpha}$,
\begin{eqnarray}
H_{1} = \frac{1}{2} \int d^2 r 
    (\Lambda _1 \delta \phi _{1} + \Lambda _2 \delta \phi _{2})(\delta l)^2
\end{eqnarray}
is the nonlinear coupling between $\delta l$ and $\delta \phi _{\alpha}$,
and $H_{\phi}$ includes terms which are linear in $\delta \phi _{\alpha}$.

First I discuss the contribution from $H_0$, i.e.,
the Gaussian fluctuations around the hard
membrane solution.  In this Gaussian approximation, $F_c$ has acquired two 
correction terms which can be expressed by 
\begin{eqnarray}
- \frac{1}{2} \sum _q \ln \frac{2 \pi }{q^4 + \Gamma q^2 + m_+}
- \sum _q 
    \frac{| l_M \delta m_+({\bf q})+ \triangle _h \delta m_{-}({\bf q})|^2}
         {q^4 + \Gamma q^2 + m_+},
\label{eq:momentumG}
\end{eqnarray}  
for convenience I have defined
$m _{\pm}=\Lambda _1 \phi _{1M} \pm \Lambda _2 \phi _{2M}$,
and $\delta m_{\pm} 
   = \Lambda _1 \delta \phi _{1} \pm \Lambda _2 \delta \phi _{2}$.
The first term is independent of $\delta \phi _{\alpha}$, therefore I neglect
it in the rest of the discussion.  The second term is a
membrane-mediated nonlocal interaction between the junctions.  This interaction
has two characteristic lengths: $m_+^{-1/4}$ is the distance it takes for
a perturbation in membrane-membrane distance to relax back to its hard membrane
solution due to the membrane bending rigidity, another length is 
$\Gamma ^{-1/2}$, for lengths greater than
$\Gamma^{-1/2}$ the elasticity of the membrane is dominated by the surface
tension of the membrane, and the contribution from the bending rigidity is 
negligible.  In the rest of this article, I focus on the case when $\Gamma 
< \sqrt{m_+}$, in which the membrane bending rigidity is the dominant effect
which drives a perturbation in $h$ back to $h_M$, thus the contribution
from surface tension of the membranes is negligible.    
To understand the nature of the nonlocal interaction, 
it is convenient to transform the second term to real space. 
Calculations in the Appendix show that, when $\Gamma < \sqrt{m_+}$, 
the second term in the real space has the form which is derived in 
Eq.~(\ref{eq:FG})
\begin{eqnarray}
-\int d^2 r \int d^2 r' \frac{\triangle _h^2}{8\pi \sqrt{m_+}} 
   G(|{\bf r}-{\bf r'}| m_+^{1/4}) 
 \left[ \left(1-\frac{m_-}{m_+} \right) \Lambda _1 \delta \phi _1 ({\bf r})
       -\left(1+\frac{m_-}{m_+} \right) \Lambda _2 \delta \phi _2 ({\bf r})
 \right]  \nonumber \\
\times  
 \left[ \left(1-\frac{m_-}{m_+} \right) \Lambda _1 \delta \phi _1 ({\bf r'})
       -\left(1+\frac{m_-}{m_+} \right) \Lambda _2 \delta \phi _2 ({\bf r'})
 \right],
\label{eq:realG}
\end{eqnarray}
where $G(x)$ is a MeijerG function~\cite{ref:mathematica}. 
$G(x)$ is vanishingly small for $x \geq 5$.  
Eq.~(\ref{eq:realG}) shows that
this membrane-mediated interaction is attractive between junctions of the same
type, and repulsive between junctions of different types.  This interaction
is short-ranged with a characteristic length $m_+^{-1/4}$.  Also notice that 
the contribution from $H_1$ is proportional to $\triangle _h^2$, i.e., 
there is no membrane-mediated interactions in the level of Gaussian 
approximations when the junctions have the same height.
The physical picture of this interaction can be seen from Fig.~5.  
A small perturbation of the junction density from the hard membrane
solution induces a deviation of membrane-membrane distance from $h_M$, 
and there is a membrane bending energy associated
with any given distribution of non-uniform membrane-membrane distance.  
To reduce the bending energy, a region with positive $\delta \phi _{1(2)}$
attracts a region with positive $\delta \phi _{1(2)}$ in order to reduce 
the elastic energy cost of a ``pit'' or a ``bump'' between these two regions 
due to the non-uniform $h({\bf r})$.  
Similarly,  a region with positive $\delta \phi _{1}$ repels a
region with positive $\delta \phi _2$, in order to reduce the bending energy
cost due to the high curvature configuration between these two regions.
This also explains the fact that these interactions vanish when both
types of junctions have the same height, i.e., $\triangle _h =0$.  A
similar kind of membrane-mediated nonlocal interaction between the junctions
is discussed in the celebrated article by Bruinsma, Goulian, and 
Pincus~\cite{ref:BPG_93}, where in their ``van der Waals regime'',
the competition between the 
potential minimum due to the van der Waals interaction between the membranes 
and another potential minimum due to the stiff membrane junctions results in
a strong interaction between the junctions.  Although there is only one
type of junctions in the system discussed in Ref.~\cite{ref:BPG_93}, the
interaction between the junctions in Ref.~\cite{ref:BPG_93} and the 
present case share the same physical mechanism, i.e., the bending 
elasticity of the membranes mediates this interaction.

Another type of nonlocal interactions between the junctions can
be studied by considering
the effect of nonlinear couplings between $\delta l$ and 
$\delta \phi _{\alpha}$. This is done by including the contributions 
form $H_1$ perturbatively to one loop order. The resulting effective
interaction free energy between the junctions, $F_c$, now has the form
\begin{eqnarray}
   F_c = F_{M} + F _{G} + F_{loop} + H_{\phi},
\label{eq:Fc-loop}
\end{eqnarray}
where $F_{M}$ is the hard membrane solution of $F_c$, $F_G$ is the contribution
from terms which are bilinear in $\delta l$ and $\delta \phi _{\alpha}$, 
and $F_{loop}$ is the contribution from the nonlinear couplings between
$\delta l$ and $\delta \phi _{\alpha}$ to one loop order. The details of
the calculations for $F_{loop}$ is discussed in the Appendix.
When $\Gamma < \sqrt{m_+}$, the result
(up to terms quadratic in $\delta \phi _{\alpha}$) is provided by 
Eq.~(\ref{eq:loop-0}), Eq.~(\ref{eq:loop-1}), and Eq.~(\ref{eq:loop-2}),
\begin{eqnarray}
F_{loop} &=& \frac{1}{16 \sqrt{m_{+}}} \int d^2 r 
           ( \Lambda _1 \delta \phi _1 + \Lambda _2 \delta \phi _2)
\nonumber \\
         && - \frac{1}{16 \sqrt{m_{+}}} \int \frac{d^2 q}{(2 \pi)^2}
            \frac{1}{q^4 + 4 m_+} 
            | \Lambda _1 \delta \phi _1(q)+ \Lambda _2 \delta \phi _2(q)|^2. 
\label{eq:Floop-momentum}
\end{eqnarray}
Here the first term is a ``renormalization'' of the chemical potentials
of the junctions due to membrane fluctuations.  This term effectively reduces
the binding energies of the junctions.  The fact that the renormalization
of the chemical potential for the softer junctions is less significant 
compared to that for the stiffer junctions is because the membrane 
fluctuations are energetically less costly for the softer junctions.
The second term is a fluctuation-induced nonlocal
interaction between the junctions, and higher order terms are neglected.
Notice that, as discussed in the Appendix, the second term in 
Eq.~(\ref{eq:Floop-momentum}) is actually an approximate form of the 
much more complicated true result, 
it provides the correct asymptotic behavior of the true result
at large and small $q$ limits in the case when $\Gamma < \sqrt{m_+}$. 
Similar to the case of
Gaussian approximation, when the fluctuation-induced interaction between the
junctions is expressed in real space, one finds that the interaction between
the junctions is nonlocal, short-ranged, and has a characteristic length on the
order of $m_+^{-1/4}$.  
Since $F_{loop}$ is non-vanishing even when $\triangle _h =0$, it is clear
that the thermal fluctuations of the membrane-membrane distance is the 
mechanism which induces the nonlocal interactions between the junctions in 
$F_{loop}$.  This is similar but not the same as the interaction between the 
junctions in the ``Helfrich regime'' discussed in Ref.~\cite{ref:BPG_93}.  
In Ref.~\cite{ref:BPG_93},
the interaction between the junctions in the Helfrich regime comes from the
collisions between the membranes.  Here in the one-loop calculation
the interaction between the junctions
comes from the fluctuations of the membrane-membrane distance
around the hard membrane solution, the effect of membrane
collisions is not included.  

When the fluctuations around the hard membrane solution are taken into 
account to one-loop order, the total free energy of the system to
second order in $\delta \phi _{\alpha}$ can be expressed by 
\begin{eqnarray}
F =     F_M 
   &+& \int \frac{d^2q}{(2 \pi)^2} \left( 
       \sum _{\alpha=1}^{2} 
           \frac{1}{2 \phi _{\alpha M}}| \delta \phi _{\alpha}(q)|^2
                                   \right) \nonumber \\
   &-& \int \frac{d^2q}{(2 \pi)^2}  \frac{\triangle _h^2}{q^4 + m_+}
       \left| \left(1-\frac{m_-}{m_+} \right) \Lambda _1 \delta \phi _1(q)
        -\left(1+\frac{m_-}{m_+} \right) \Lambda _2 \delta \phi _2(q)
	\right|^2
\nonumber \\
   &-& \frac{1}{16 \sqrt{m_{+}}} \int \frac{d^2 q}{(2 \pi)^2}
            \frac{1}{q^4 + 4 m_+} 
            | \Lambda _1 \delta \phi _1(q)+ \Lambda _2 \delta \phi _2(q)|^2
\nonumber \\
   &+& \frac{1}{16 \sqrt{m_{+}}} \int d^2 r 
           ( \Lambda _1\delta \phi _1 + \Lambda _2\delta \phi _2).
\label{eq:total}
\end{eqnarray}
Here the first term on the right hand side is the hard membrane solution, 
the second term comes from the entropy of the
junctions, the third term is the nonlocal interaction between the junctions
due to Gaussian fluctuations, the fourth and the fifth terms come from
the nonlinear couplings between $\delta l$ and $\delta \phi _{\alpha}$.
Notice that the contribution from $H_{\phi}$ does not appear in the total
free energy of the system, it cancels with
the linear terms in the expansion of the entropy of the junctions.  This is 
because $\phi _{\alpha M}$ minimizes the hard-membrane free energy, therefore
in the expansion around the hard-membrane solution
terms which are linear in $\delta \phi $ cancel with each other.  
The contribution from one-loop calculation, however, includes
terms which are linear in $\delta \phi _{\alpha}$ because they come from
the nonlinear couplings between $\delta \phi _{\alpha}$ and $\delta l$.  
An important consequence of the presence of these terms is that in general 
the equilibrium values of $\delta \phi _1$ and $\delta \phi _2$ are nonzero
due to the membrane fluctuations.  Therefore when a phase separation occurs,  
the values of $\phi _1 + \phi _2$ are different in domains
with different values of $\phi$. 

To discuss the correction of $\phi _{\alpha}$ and $h({\bf r})$ due to
the thermally activated fluctuations, it is convenient to 
express Eq.~(\ref{eq:total}) as
\begin{eqnarray}
F = F_M + \int d^2r \left( \delta \mu _1 \delta \phi _1({\bf r})
                          +\delta \mu _2 \delta \phi _2({\bf r}) \right)
   + \sum _{\alpha \beta} \int \frac{d^2q}{(2\pi)^2} M_{\alpha \beta}(q)
         \delta \phi _{\alpha}(q) \delta \phi _{\beta}(q),
\end{eqnarray}
where
\begin{eqnarray}
    \delta \mu _{\alpha} 
&=& \frac{\Lambda _{\alpha}}{16\sqrt{m_+}}, \nonumber \\
    M_{11}(q)
&=& \frac{1}{2\phi _{1M}}
 -\frac{\triangle _h^2}{q^4 + m_+}\left(1-\frac{m_-}{m_+}\right)^2 \Lambda _1^2
 -\frac{1}{16\sqrt{m_+}}\frac{\Lambda _1^2}{q^4+4m_+}, \nonumber \\
    M_{22}(q)
&=& \frac{1}{2\phi _{2M}}
 -\frac{\triangle _h^2}{q^4 + m_+}\left(1+\frac{m_-}{m_+}\right)^2 \Lambda _2^2
 -\frac{1}{16\sqrt{m_+}}\frac{\Lambda _2^2}{q^4+4m_+}, \nonumber \\
    M_{12}(q) &=& M_{21}(q) 
= \frac{\triangle _h^2}{q^4 + m_+}\left(1-\left(\frac{m_-}{m_+}\right)^2\right)
  \Lambda _1 \Lambda _2
-\frac{1}{16\sqrt{m_+}}\frac{\Lambda _1 \Lambda _2}{q^4+4m_+}.
\end{eqnarray}
Now $\delta \phi _{\alpha}(q)$ can be expressed by $\delta \mu _{\alpha}$
and $M_{\alpha \beta}$,
\begin{eqnarray}
\delta \phi _1(q) + \delta \phi _2(q) &=&
\delta (q)\times \frac{-1}{2 \det{\bf M}} 
\left\{ \left( M_{22}-M_{21} \right) \delta \mu _1
       +\left( M_{11}-M_{12} \right) \delta \mu _2 
\right\}, \nonumber \\
\delta \phi _1(q) - \delta \phi _2(q) &=&
\delta (q)\times \frac{-1}{2 \det{\bf M}} 
\left\{ \left( M_{22}+M_{21} \right) \delta \mu _1
       -\left( M_{11}+M_{12} \right) \delta \mu _2 
\right\},
\label{eq:deltaphi}
\end{eqnarray}  
where $\det {\bf M}= M_{11}M_{22}-M_{12}M_{21}$.
This rather complicated expression shows that, besides
$\mu _-$, $\lambda$, and $\triangle _h^2 \Lambda _+$, 
the answer to the question of which domain has higher total junction density 
when the phase coexistence occurs also depends on the values of 
$\triangle _h^2$ 
and $\phi _{\alpha M}$ (to determine $\phi _{\alpha M}$, one needs to know
the value of $\mu _+$)!  
In this article I shall not discuss the details of the values of
$\phi _1 + \phi _2$ for different given parameters in the theory, 
but simply comment that when $\det {\bf M}$ is positive, the phase diagram
of the hard membrane solution is not modified by the thermal fluctuations.
However, when  $\det {\bf M} <0$, the hard membrane solution is not stable 
at any finite temperature.  
The result in Eq.~(\ref{eq:deltaphi}) also provides some 
criteria for the current analysis.  For example,
when the fluctuations are large,
the deviation from hard membrane solution can no longer be treated by 
perturbation theory.  This is true when 
$\delta \phi _{\alpha}/ \phi _{\alpha} \sim \mathcal{O}(1)$.  Since
$\delta \phi _{\alpha}$ becomes large for small $\det {\bf M}$, which
occurs at small $m_+ =\Lambda _1 \phi _{1M} + \Lambda _2 \phi _{2M}$,
I conclude that the perturbation theory breaks down at small junction
densities. Finally, I point out that
the collisions between the membranes are also neglected in the present 
analysis, this approximation is valid when the fluctuations of 
membrane-membrane distance is not large, i.e., when 
\begin{eqnarray}
\frac{\sqrt{<(\delta l)^2>_0}}{h_M}
= \left( \int \frac{d^2q}{(2\pi)^2} \frac{1}{q^4 + m_+}\right)^{1/2}
  \frac{1}{h_M}
\approx \frac{1}{4 m_+^{1/4} h_M} 
\leq \mathcal{O}(1).
\end{eqnarray}
In the regime where 
$m_+^{1/4} h_M =   (\Lambda _1 \phi _{1M} + \Lambda _2 \phi _{2M})^{1/4} h_M
              \leq \mathcal{O}(1)$, 
i.e., when the junction densities are small, or the when junctions are 
very soft, or when the junctions are very ``short'', 
the contributions from membrane collisions should be
taken into account for a complete analysis of this system. 
Thus, when the membrane fluctuations or the membrane collisions become 
important, numerical simulations~\cite{ref:chen_future} or other methods which
take the full membrane fluctuations into account should be applied to study 
the physics of this system.

\section{Summary}
\label{sec:summary}
  I have discussed the phase separation of 
multiple species membrane junctions 
induced by membrane-membrane adhesion with a continuum theory.  
In the hard membrane approximation where
the membrane-membrane distance and junction densities are assumed to be
constants, we find that $\triangle _h^2 \Lambda _{+}$ and $\lambda $ are the
important parameters which governs the junction separation.  When $\lambda =0$,
both types of junctions have the same rigidity, and the junction separation
is driven by the height difference of the junctions.  Under this
condition the junction separation is very similar to the Flory-Huggins theory
for a binary mixture. 
Phase separation occurs when $\triangle _h ^2 \Lambda _{+} >2$ and $\mu _-=0$, 
the phase coexistence curve ends at a  critical point 
$\mu _{-}=0$, $\triangle _h ^2 \Lambda _{+}=2$.  When $\lambda \neq 0$, the
junctions have different rigidities, and the softer junctions are 
easier to form than the stiffer junctions.  Therefore the softer junctions
have a tendency to aggregate.
In this more general case, the height difference and the junction rigidities
difference both drive the phase separation, thus the phase 
separation can occur at  $\triangle _h ^2 \Lambda _{+}<2$.

  The Gaussian fluctuations around the hard membrane solution reveals
a membrane-mediated nonlocal interaction between the junctions.  This
interaction is short-ranged, which decays with a characteristic length 
$(\Lambda _+ \phi _{+M} + \Lambda _- \phi _{-M})^{-1/4}$, it is 
attractive between the same type of junctions, but repulsive
between different types of junctions. 
The strength of this interaction is proportional to $\triangle _h ^2$,
and it is due to the membrane bending energy cost between
regions with different junction densities.
Perturbation theory to one-loop order shows other effects of thermal 
fluctuations, this includes a renormalization of the chemical potential of the 
junctions which effectively reduces the binding energies of the junctions, 
and a nonlocal interaction between the junctions which is 
independent of $\triangle _h$.  The fact that the contribution from one-loop 
calculation is non-vanishing even when junctions of type-1 and type-2 have 
the same height indicates that this contribution is a result of thermal
fluctuations of the membranes.  
Hence it is non-vanishing at all finite temperatures.
When the contribution from one-loop calculation becomes very large, the hard
membrane solution is qualitatively incorrect, and the effects of thermal
fluctuations is a dominant factor.  This can occur at very low junction
densities.  The Gaussian fluctuations of the
membrane-membrane distance also provide another limit of the present analysis:
the mean squared fluctuations of the membrane-membrane distance should
be small compare to $h_M$.  As a result, the analysis in this article does 
not provide the complete physical picture of the system for very soft
or very short junctions, either.

  In summary, mean field and fluctuation analysis of a simple coarse grained 
model for adhesion-induced phase separation of multiple species of 
membrane junctions is studied in this article.  
This model shows rich behaviors which capture much of the physics of 
multi-species membrane junction separation induced by adhesion.  
I show that not only the difference
of junction height, but also the difference of junction rigidities, and the
membrane-mediated interactions between the junctions play important roles in 
the junction separation.  The fluctuation analysis also shows that current
analysis does not provide the complete physical picture  for systems with 
very soft or very short junctions, or in the situation when the
junction densities are extremely low, where the thermally activated membrane
fluctuations or the Helfrich repulsion between
the membranes become important interactions in the system~\cite{ref:BPG_93}.  
In this regime, numerical 
simulations~\cite{ref:lipowsky_PRE_01,ref:chen_future} should provide 
valuable information on the distribution of the junctions, as well as a 
complete picture of the phase diagram which includes the bind/unbinding
transition between the membranes, and adhesion-induced junction separations.

\section*{Acknowledgment}
   It is a pleasure to thank David Lu and David Jasnow 
for very helpful discussions.  
This work is supported by the National Science Council of 
the Republic of China (Taiwan) under grant No. NSC 90-2112-M-008-053.   

\section*{Appendix}
   In this Appendix, I discuss the details of some calculations mentioned in 
the text.  For simplicity I define 
\begin{eqnarray}
  \delta m_+ ({\bf r}) 
&=& \Lambda _1 \delta \phi _1({\bf r}) + \Lambda _2 \delta \phi _2({\bf r})
\nonumber \\
 \delta m_- ({\bf r}) 
&=& \Lambda _1 \delta \phi _1({\bf r}) - \Lambda _2 \delta \phi _2({\bf r}).
\end{eqnarray}
First, an integral which is very useful for the rest of this
Appendix is calculated. 
\begin{eqnarray}
\int \frac{d^2q}{(2 \pi)^2} \frac{A(q) B(-q)}{q^4 + m_+}
&=&\int d^2 r \int d^2 r' \int \frac{d^2q}{(2 \pi)^2}
\frac{A({\bf r}) B({\bf r'})e^{i {\bf q} \cdot ({\bf r}-{\bf r'})}}{q^4 + m_+}
\nonumber \\
&=& \int d^2 r \int d^2 r'
    \frac{1}{\sqrt{m_+}} \int _0 ^{\infty} \frac{dx}{2 \pi} 
    \frac{x J_0(x|{\bf r}-{\bf r'}|m_+^{1/4})}{x^4 +1}
    A({\bf r}) B({\bf r'})  \nonumber \\
&=& \frac{1}{8\pi \sqrt{m_+}} \int d^2 r \int d^2 r'
    G(|{\bf r}-{\bf r'}|m_+^{1/4}) A({\bf r})B({\bf r'}),
\label{eq:integral}
\end{eqnarray}
where $G(x)$ is a MeijerG function~\cite{ref:mathematica}.  Also,
\begin{eqnarray}
G(x) \approx \left\{ \begin{array}{ll}
                     \pi , & x \ll 1 \nonumber \\
                      0 ,  & x \geq 5.
                     \end{array}
		\right.
\end{eqnarray}
The shape of $G(x)$ is plotted in Fig.~6.

Now I consider the nonlocal interaction between the junctions
in the Gaussian approximation.  Neglecting the first term of 
Eq.~(\ref{eq:momentumG}), the second term can be expressed by 
\begin{eqnarray}
 F_G &=& -\sum _q 
    \frac{| l_M \delta m_+({\bf q})+ \triangle _h \delta m_{-}({\bf q})|^2}
         {q^4 + \Gamma q^2 + m_+} \nonumber \\
    &=& -\int \frac{d^2 q}{(2 \pi)^2} 
\frac{| l_M \delta m_+({\bf q})+ \triangle _h \delta m_{-}({\bf q})|^2}
         {q^4 + \Gamma q^2 + m_+} \nonumber \\
    &\approx& -\int d^2 r \int d^2 r' \frac{\triangle _h^2}{8\pi \sqrt{m_+}} 
   G(|{\bf r}-{\bf r'}| m_+^{1/4}) 
 \left[ \left(1-\frac{m_-}{m_+} \right) \Lambda _1 \delta \phi _1 ({\bf r})
       -\left(1+\frac{m_-}{m_+} \right) \Lambda _2 \delta \phi _2 ({\bf r})
 \right]  \nonumber \\
&&\times  
 \left[ \left(1-\frac{m_-}{m_+} \right) \Lambda _1 \delta \phi _1 ({\bf r'})
       -\left(1+\frac{m_-}{m_+} \right) \Lambda _2 \delta \phi _2 ({\bf r'})
 \right],
\label{eq:FG}
\end{eqnarray}
where the last expression holds when $\Gamma < \sqrt{m_+}$, and 
the integral in Eq.~(\ref{eq:integral}) is used to calculate the Fourier 
transformation from the momentum space to the real
space.  The range of this membrane mediated interaction between the junctions
is determined by the shape of $G(x)$, which sets the length scale of
this interaction to $m_+^{-1/4}$.

Next I show that the interaction between the junctions due to the contribution
of $H_1$ is of the form in Eq.~(\ref{eq:Floop-momentum}).  The effective
interaction free energy between the junctions is 
\begin{eqnarray}
F_c &=& - \ln \left[ \int D[h] e^{-H_M-H_0-H_1-H_{\phi} } \right] \nonumber \\
    &=& H_M +H_{\phi}- \ln \left[  \int D[\delta l] e^{-H_0-H_1} \right].
\end{eqnarray}
When the contribution of $H_1$ is included by a one-loop calculation, 
\begin{eqnarray}
F_c &=&   H_M + H_{\phi} + F_G + <H_1>_0 
     - \frac{1}{2} \left( <H_1^2>_0 - <H_1>_0^2 \right) \nonumber \\
    &\equiv&   F_M+ F_G +F_{loop}+H_{\phi}.
\label{eq:loop-0}
\end{eqnarray}
Here 
\begin{eqnarray}
F_M=H_M, \ \ \  F_G=-\ln[ \int D[\delta l] e^{-H_0}], \ \ \ \mbox{and} \
<\mathcal{O}>_0 = \frac{\int D[\delta l]\ 
 \mathcal{O} e^{-H_0}}{ \int D[\delta l] e^{-H_0}} 
\end{eqnarray}
for any $\mathcal{O}$.
The calculation of $<H_1>_0$ is straightforward, which in the case
$\Gamma < \sqrt{m_+}$ leads to 
\begin{eqnarray}
<H_1>_0&=&\frac{1}{2} 
        \int d^2 r (\Lambda _1 \delta \phi _1({\bf r}) 
                  + \Lambda _2 \delta \phi _2({\bf r}))
        < (\delta l({\bf r}) )^2>_0  \nonumber \\
       &=&\frac{1}{2}
        \left( \int \frac{d^2 q}{(2\pi)^2} \frac{1}{q^4 + \Gamma q^2 + m_+}
        \right) (\Lambda _1 \delta \phi _1({\bf r})
               + \Lambda _2 \delta \phi _2({\bf r}) )
\nonumber \\
       &\approx & \frac{1}{16 \sqrt{m_+}} \int d^2 r 
          (\Lambda _1 \delta \phi _1({\bf r}) 
         + \Lambda _2 \delta \phi _2({\bf r})).
\label{eq:loop-1}
\end{eqnarray}
Terms of higher order than $\delta \phi _{\alpha}^2$ have been neglected.
The calculation for $<H_1^2>_0$ is longer but also straightforward,
\begin{eqnarray}
<H_1^2>_0&=&
 \frac{1}{4} \int d^2 r \int d^2 r' 
 <\delta l({\bf r}) \delta l({\bf r}) \delta l({\bf r'}) \delta l({\bf r'}) >_0
 \delta m_+({\bf r}) \delta m_+({\bf r'})
\nonumber \\
&=&\frac{1}{2} \int \frac{d^2q}{(2\pi)^2} \int \frac{d^2q'}{(2\pi)^2}
\frac{1}{q'^4 + \Gamma q'^2 + m_+} 
\frac{1}{({\bf q}+{\bf q'})^4 + \Gamma ({\bf q}+{\bf q'})^2 + m_+}
| \delta m_+({\bf q}) |^2 
+ <H_1>_0^2.
\nonumber \\
\end{eqnarray}
Again, terms of higher order than $\delta \phi _{\alpha}^2$ are neglected.
When the contribution from the surface tension can be neglected, the 
following form provides a good approximation of
$<H_1^2>_0 - <H_1>_0^2$,~\cite{ref:netz_PRE_95} this form gives the correct
asymptotic behavior of the true result at large and small $q$ limits,
\begin{eqnarray}
<H_1^2>_0 - <H_1>_0^2 \ 
&\approx&
\frac{1}{8 \sqrt{m_+}} \int d^2 q \frac{|\delta m_+(q)|^2}{q^4 + 4 m_+} 
\nonumber \\
&=&\frac{1}{64\pi\sqrt{m_+^3}} \int d^2 r \int d^2 r'
G(|{\bf r}-{\bf r'}| (4m_+)^{1/4})
\left( \Lambda _1 \delta \phi _1({\bf r}) + \Lambda _2 \delta \phi _2({\bf r})
\right) 
\nonumber \\
&& \times
\left( \Lambda _1 \delta \phi _1({\bf r'})+ \Lambda _2 \delta \phi _2({\bf r'})
\right) 
\label{eq:loop-2}
\end{eqnarray}
Putting $<H_1>_0$ and $<H_1^2>_0 - <H_1>_0^2$ together leads to
the resulting expression of $F_c$ in the one-loop order, which is given
in Eq.~(\ref{eq:Fc-loop}) and Eq.~(\ref{eq:Floop-momentum}).  The real
space form of $<H_1^2>_0 - <H_1>_0^2$ also shows that the membrane fluctuation
induced interaction between the junctions is short ranged with a 
characteristic length on the order of $m_+^{-1/4}$

\newpage
{\bf Figure Captions}
\begin{itemize}
\begin{figure}[h]
\epsfxsize= 3.5 in
\epsfbox{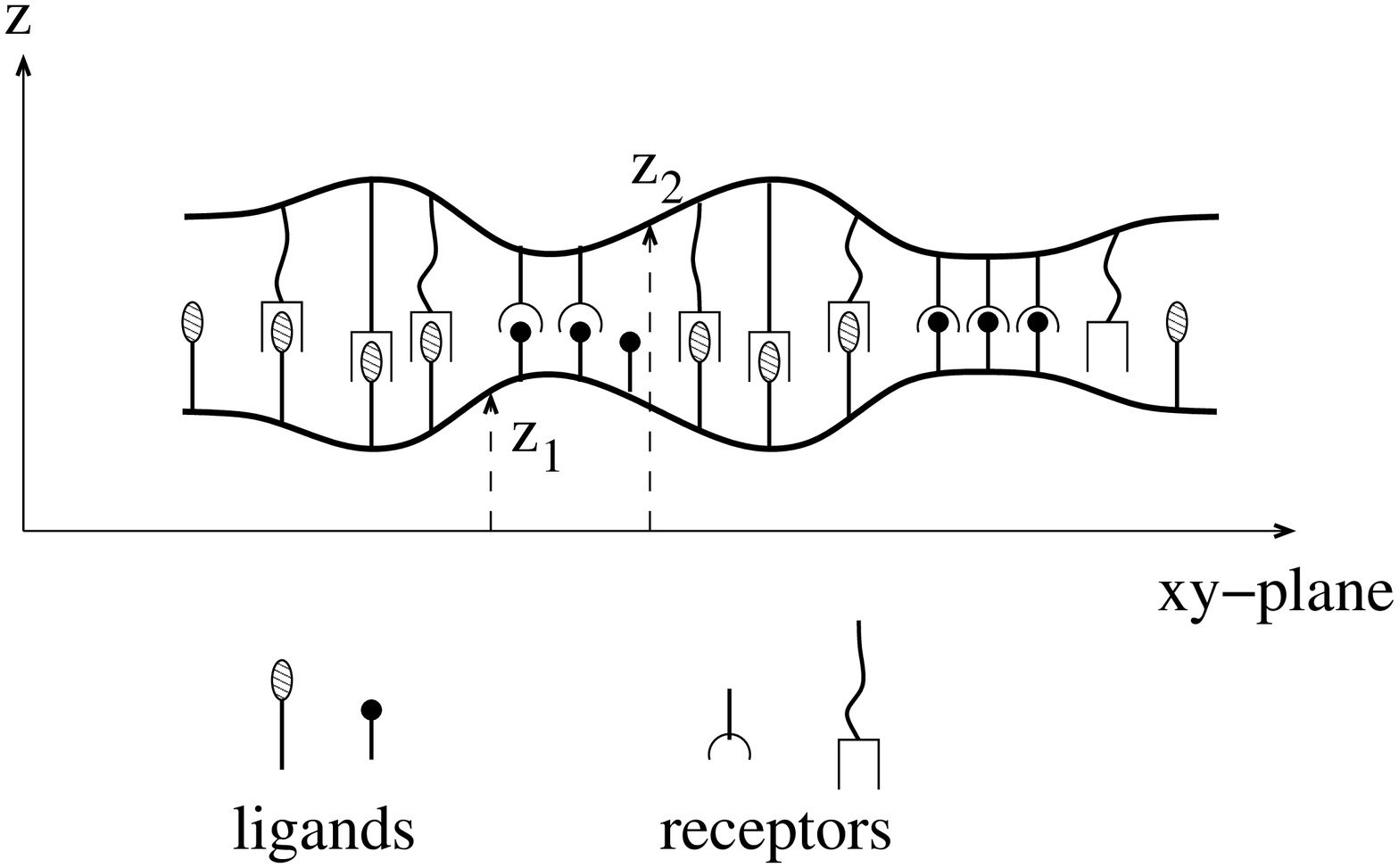}
\end{figure}
\item{Fig.1  Schematics of the system.  The membrane heights are 
    $z_1({\bf r})$ and $z_2({\bf r})$ from the reference plane.
    There are two types of receptors in one membrane, two types of ligands
    in another membrane.
    Two types of junctions can be formed from the ligands and receptors. 
    They have different natural lengths $h_1$ and $h_2$.  
    In general, different types of junctions
    also have different rigidities.  The softer junctions are easier to be
    stretched or compressed from their natural length.
\label{fig:config}}
\begin{figure}[h]
\epsfxsize= 3.5 in
\epsfbox{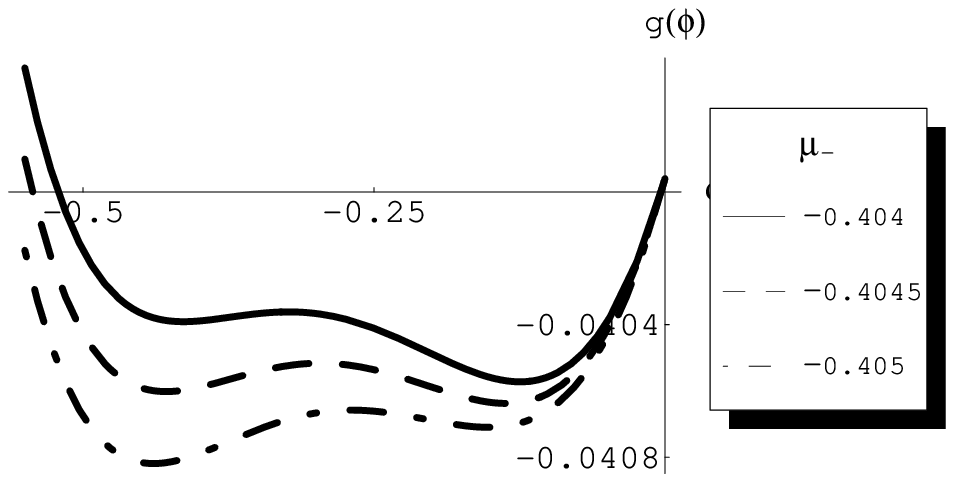}
\end{figure}
\item{Fig.2  The shape of 
    $g(\phi)$ with different values of $\mu _{-}$ when 
    $\lambda=0.2$, and $\triangle _h^2 \Lambda _{+} = 1.998$. 
    Solid line: $\mu = -0.404$, dashed line: $\mu =-0.4045$, dash-dotted line:
    $\mu = -0.405$. Phase coexistence occurs at $\mu \approx -0.4045$ even
    though  $\triangle _h^2 \Lambda _{+} <2.0$. 
\label{fig:enhance}}
\newpage
\begin{figure}[h]
\epsfxsize= 3.5 in
\epsfbox{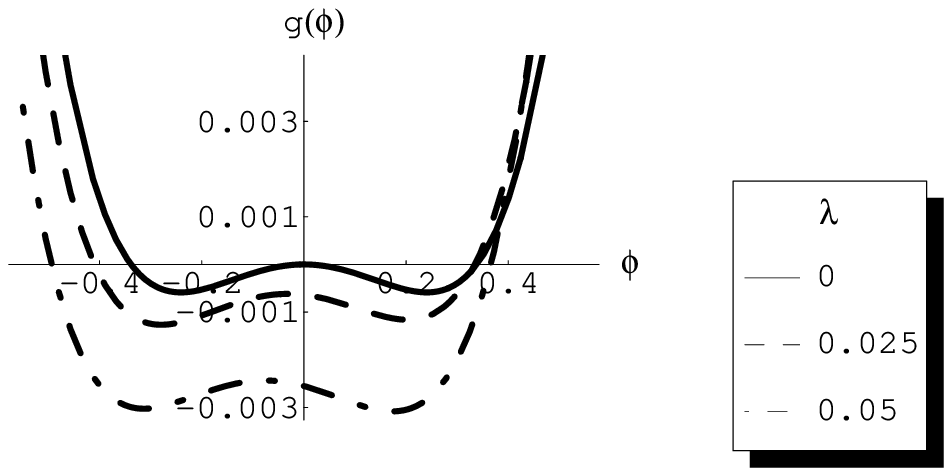}
\end{figure}
\item{Fig.3 $g(\phi)$ in the phase coexistence
    for different values of $\lambda$ with 
    $\triangle _h^2 \Lambda _{+}=2.04$. 
    Solid line: $\lambda =0$, dashed line: $\lambda =0.025$, dash-dotted line:
    $\lambda =0.05$.  $\lambda =0$ curve is symmetric around $\phi =0$, 
    $\lambda >0$ curves shows that the positions of the minima are shifted 
    towards smaller $\phi$ values, i.e., 
    softer junctions are the favored species.
\label{fig:asymmetry} }

\begin{figure}[h]
\epsfxsize= 3 in
\epsfbox{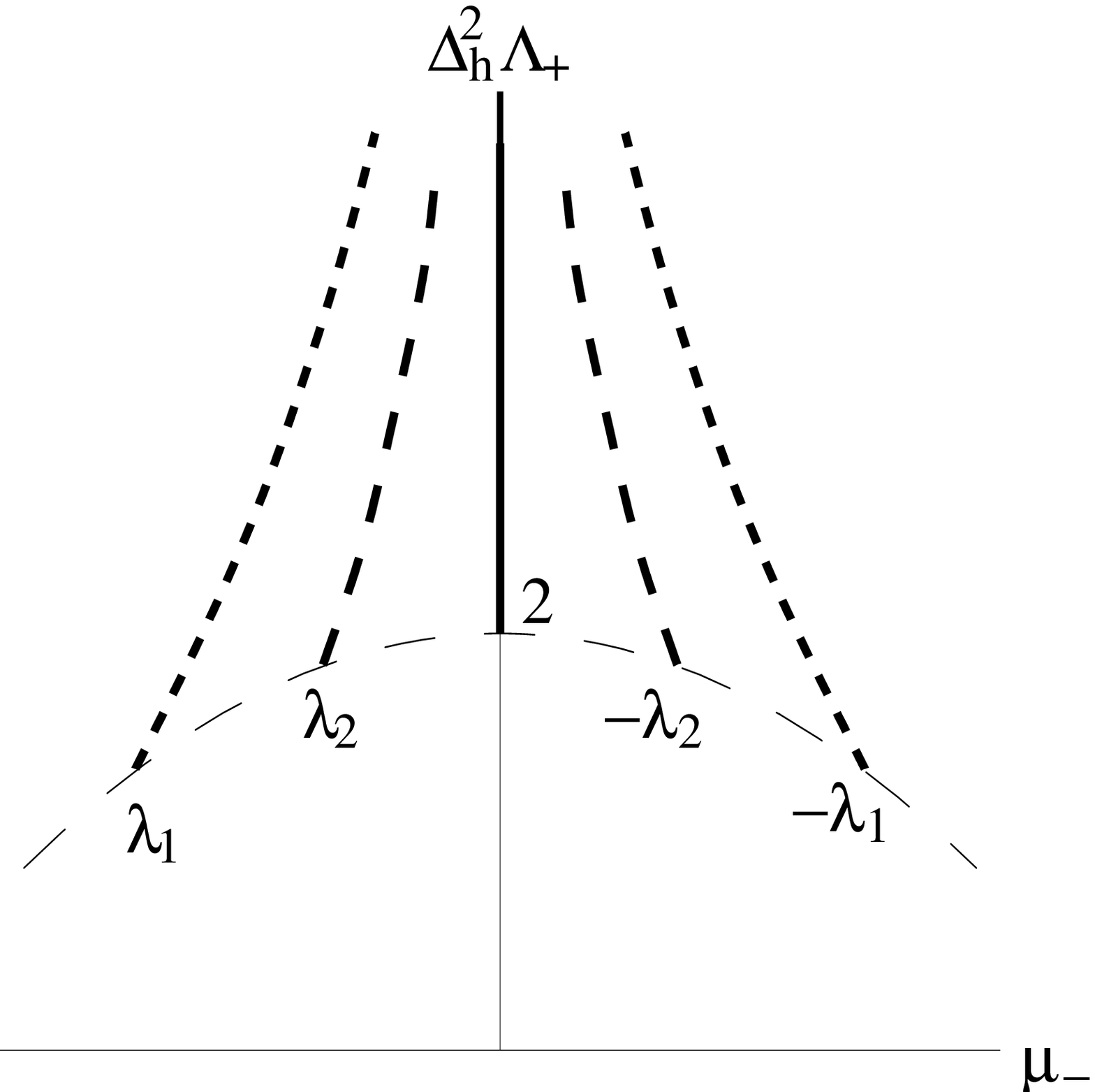}
\end{figure} 
\item{Fig.4 Schematics of the phase coexistence curves near 
 $\triangle _h^2 \Lambda _{+}=2$ for $\lambda \ll1$ for $\lambda =0$
(thick solid line), $\lambda = \pm \lambda _1$ (thick short-dashed lines),
and $\lambda = \pm \lambda _2$ (thick long-dashed lines).   
 $\lambda _1 > \lambda _2 >0$.  
The thin dashed curve is the position of the end points of the phase 
coexistence curves, this is given by 
$\triangle _h^2  \Lambda _{+}\approx 2(1-9\lambda^2/4), \ 
    \mu _{-}\approx -2\lambda$.
The curves move towards the $\lambda=0$ phase boundary as 
$\triangle _h^2  \Lambda _{+}$ increases because the effect of junction height
mismatch becomes more important.  }

\newpage

\begin{figure}[h]
\epsfxsize= 3.5 in
\epsfbox{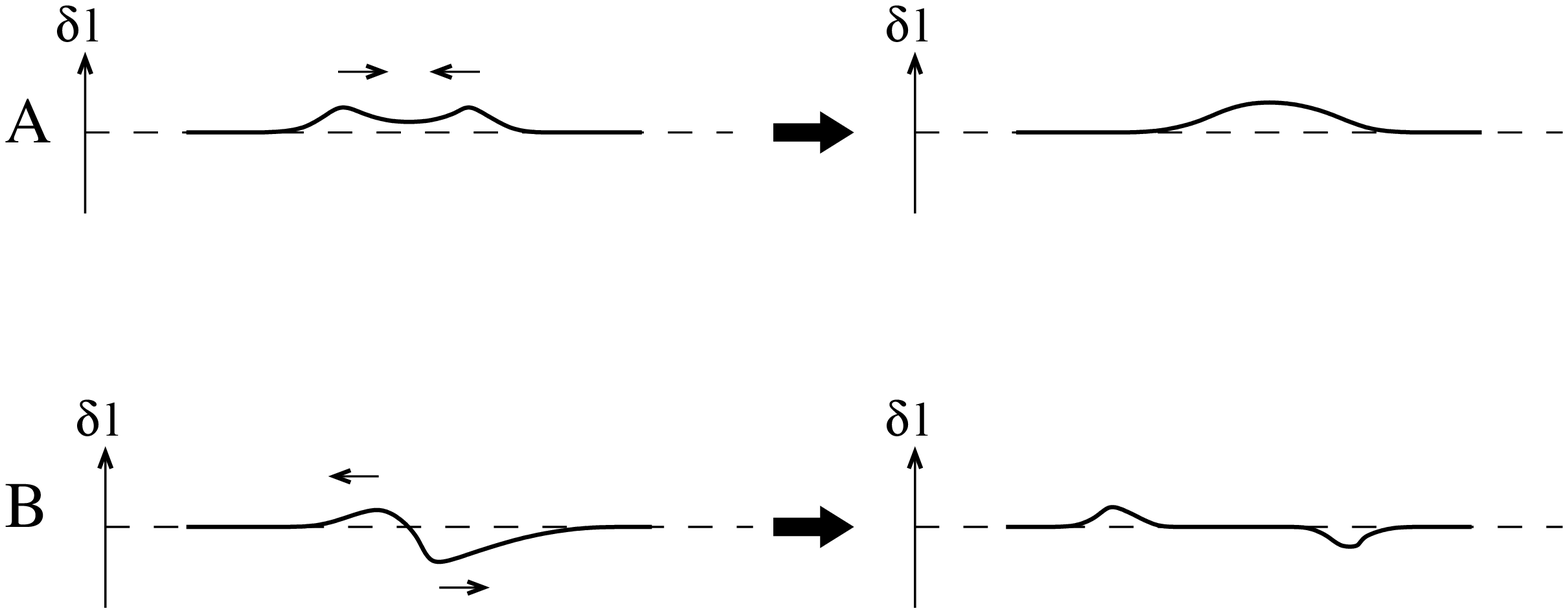}
\end{figure} 
\item{Fig.5 Membrane-mediated interaction revealed by the Gaussian 
approximation.  This interaction comes from the bilinear coupling between
$\delta l$ and $\delta \phi _{\alpha}$.  (A) A small region which has 
higher density in the junctions with greater natural height (or lower
density in the junctions with smaller natural height) induces a 
positive $\delta l$.  Two regions with positive $\delta l$ can reduce the
bending elastic energy of the membranes by moving close to each other.  
Similarly, a region with negative $\delta l$ attracts another region with
negative $\delta l$ due to the cost of membrane bending energy.  (B) A
small region with positive $\delta l$ repels with a region with negative 
$\delta l$ because of the bending elastic energy cost of the
the high curvature region between them. } 

\begin{figure}[h]
\epsfxsize= 3 in
\epsfbox{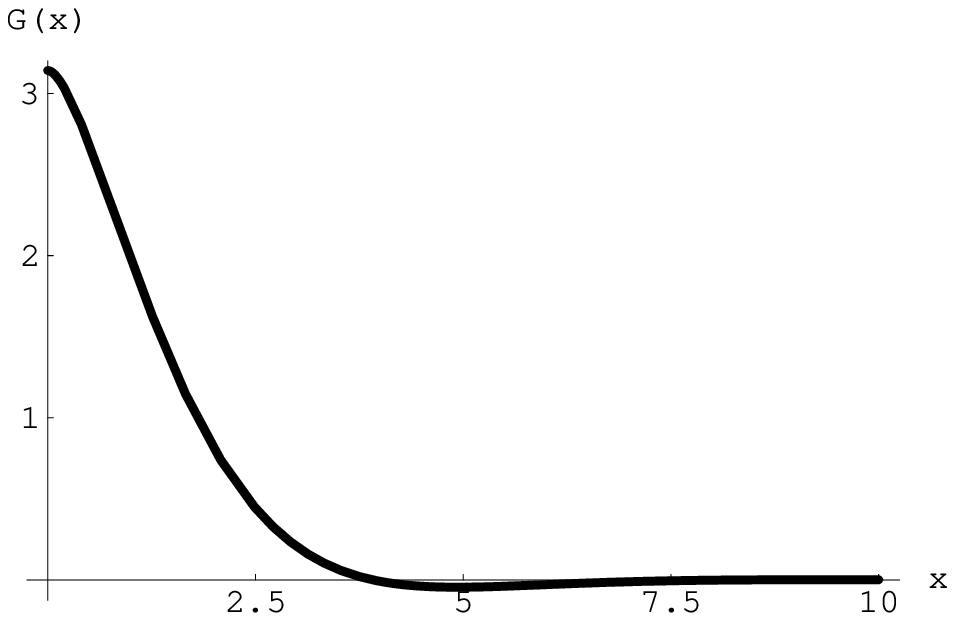}
\end{figure} 
\item{Fig.6 The shape of the MeijerG function $G(x)$.  
      Although $G(x)$ oscillates
      very weakly, and has a local minimum close to $x=5$, 
      the important feature of $G(x)$ is
      that this function is vanishingly small when $x\geq 5$.} 
\end{itemize}

\end{document}